\documentclass[aps,twocolumn,amssymb,secnumarabic,nobibnotes,superscriptaddress]{revtex4-1}

\usepackage{graphicx}% Include figure files
\usepackage{dcolumn}% Align table columns on decimal point
\usepackage{bm}% bold math
\usepackage{amsmath}
\usepackage{color}
\usepackage{ulem}
\usepackage{threeparttable}
\usepackage{enumerate}
\usepackage{paralist}

\begin{document}

\title{Na$_5$Co$_{15.5}$Te$_6$O$_{36}$: an $S$ = 1/2 stacked Ising kagome antiferromagnet with a partially disordered ground state}

\author{Z. Y. Zhao}
\email{zhiyingzhao@fjirsm.ac.cn}
\affiliation{State Key Laboratory of Structural Chemistry, Fujian Institute of Research on the Structure of Matter, Chinese Academy of Sciences, Fuzhou, Fujian 350002, People's Republic of China}

\author{X. Y. Yue}
\affiliation{Institute of Physical Science and Information Technology, Anhui University, Hefei, Anhui 230601, People's Republic of China}

\author{J. Y. Li}
\affiliation{State Key Laboratory of Structural Chemistry, Fujian Institute of Research on the Structure of Matter, Chinese Academy of Sciences, Fuzhou, Fujian 350002, People's Republic of China}

\author{N. Li}
\affiliation{Department of Physics, Hefei National Laboratory for Physical Sciences at Microscale, and Key Laboratory of Strongly-Coupled Quantum Matter Physics (CAS), University of Science and Technology of China, Hefei, Anhui 230026, People's Republic of China}

\author{H. L. Che}
\affiliation{Department of Physics, Hefei National Laboratory for Physical Sciences at Microscale, and Key Laboratory of Strongly-Coupled Quantum Matter Physics (CAS), University of Science and Technology of China, Hefei, Anhui 230026, People's Republic of China}

\author{X. F. Sun}
\affiliation{Institute of Physical Science and Information Technology, Anhui University, Hefei, Anhui 230601, People's Republic of China}
\affiliation{Department of Physics, Hefei National Laboratory for Physical Sciences at Microscale, and Key Laboratory of Strongly-Coupled Quantum Matter Physics (CAS), University of Science and Technology of China, Hefei, Anhui 230026, People's Republic of China}
%\affiliation{Collaborative Innovation Center of Advanced Microstructures, Nanjing University, Nanjing, Jiangsu 210093, People's Republic of China}

\author{Z. Z. He}
\affiliation{State Key Laboratory of Structural Chemistry, Fujian Institute of Research on the Structure of Matter, Chinese Academy of Sciences, Fuzhou, Fujian 350002, People's Republic of China}

\date{\today}

\begin{abstract}

$S$ = 1/2 Ising kagome antiferromagnets (KAFs) are rare in reality. In this work, we report the magnetic ground state and field-induced transitions of an $S$ = 1/2 Ising antiferromagnet with a stacked kagome geometry, Na$_5$Co$_{15.5}$Te$_6$O$_{36}$. Upon cooling, magnetic susceptibility measurement reveals three successive magnetic transitions at $T_1$ = 45.2 K, $T_2$ = 35.4 K, and $T_3 \sim$ 20 K. When the magnetic field is applied along the Ising $c$ axis, a strikingly anomalous initial magnetization lying outside the hysteresis loop is observed below $T_3$. Through the detailed characterizations, the magnetic field $vs$ temperature phase diagram is established. A partially disordered antiferromagnetic state is proposed below $T_1$, which becomes frozen below $T_3$. Na$_5$Co$_{15.5}$Te$_6$O$_{36}$ is therefore a promising candidate for $S$ = 1/2 Ising KAFs, and this unique composite structure may shed light on the exploratory path for $S$ = 1/2 Ising KAFs.

\end{abstract}

\maketitle

\section{Introduction}

Geometry of corner-sharing triangles that impedes the magnetic order renders the $S$ = 1/2 kagome antiferromagnets (KAFs) a recognized cornerstone to cognize the exotic quantum phenomena \cite{PhysToday-59-24,RMP-88-041002,CRP-17-455,Nature-464-199,AdvPhys-67-149,RMP-89-025003}. For Heisenberg spins on a structurally perfect kagome lattice, the degenerate manifold ground states prohibit the magnetic order at zero Kelvin \cite{PRB-56-2521}. $S$ = 1/2 Heisenberg KAF is therefore an excellent motif for realizing the fascinating quantum spin liquid (QSL), a long-range entangled state of matter hosting emergent gauge fields and fractionalized excitations \cite{Nature-464-199,RMP-89-025003,RPP-80-016502}. A number of compounds are discovered to be QSL candidates, including the celebrated Herbertsmithite ZnCu$_3$(OH)$_6$Cl$_2$ \cite{Nature-492-406} and its relatives \cite{PRB-84-100401(R),CPL-36-017502,CPL-34-077502,PRL-109-037208,NJP-16-093011}, as well as the variant breathing KAF based on V$^{4+}$ ($d^1$, $S$ = 1/2) ions \cite{NatChem-3-801,PRL-110-207208,PRL-118-237203}. On the other hand, for Ising KAFs, the highly frustrated spins can also give rise to various types of magnetic order \cite{JPAMG-21-2195,JPSJ-62-3943,PRL-59-1629}. However, as far as we know, layered $S$ = 1/2 Ising KAFs have been scarcely found in reality, and the rare candidates hinder the exploration of novel phenomena and relevant physics.

$S$ = 1/2 Ising chain, frequently observed in cobalt oxides due to the Kramers nature of Co ions, is also an attractive system and has been reported to show a variety of novel physics, for instance, quantum criticality \cite{PRL-123-067203,PRL-120-207205}, topological excitations \cite{NatPhys-14-716}, Bethe strings \cite{Nature-554-219}, and excited bound states with emergent $E_8$ symmetry \cite{Science-327-177}, etc. It has been found that a composite geometry consisted of Ising chains in a two-dimensional array can inherit the excellent traits of both geometries. For example, in the $S$ = 1/2 stacked Ising triangular lattice antiferromagnets (TAFs) $A$Co$X_3$ ($A$ = Rb, Cs; $X$ = Cl, Br) and Ca$_3$Co$_2$O$_6$, in which Ising Co chains are arranged in a triangular lattice, not only the domain-wall excitations of one-dimensional antiferromagnetic (AFM) Ising chain \cite{PRL-124-257201,NatPhys-1-159}, but also the magnetization plateau and partially disordered state in triangular lattice \cite{PRB-70-044424,PRB-103-094424,PRB-49-12299,JPSJ-77-104703} are observed. Besides, a disordered ground state is also found in Ba$_3$Co$_2$O$_6$(CO$_3$)$_{0.7}$ due to its honeycomb structure \cite{PRB-89-054431,CM-12-966}. Inspired by this, placing $S$ = 1/2 Ising chains on a kagome geometry is an alternative approach to realize the $S$ = 1/2 Ising KAFs, named stacked Ising KAFs. In addition, such a composite structure also offers a platform for chemists to search for potential $S$ = 1/2 Ising KAF candidates and uncover diverse physics.

\begin{figure}
\includegraphics[clip,width=8.5cm]{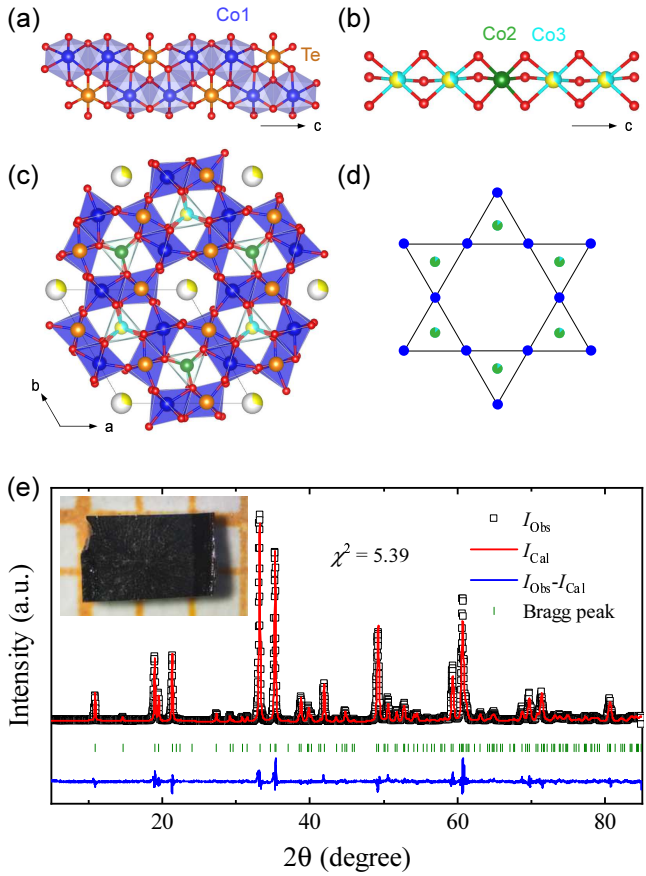}
\caption{(Color online) Crystal structure of NCTO. (a) Zig-zag chain of Co1 atoms. (b) Alternate arrangement of Co2 and Co3 atoms. Co3 site is partially occupied by Na atom. (c) Crystal structure projected in the $ab$ plane. Na atoms (yellow) sit in the hexagonal channels. The solid frame denotes the unit cell. (d) Topological spin lattice in the $ab$ plane. Co1 chains (blue symbols) build a perfect stacked kagome lattice, and Co2/Co3 chains (green/cyan symbols) locate in the center of each triangle. (e) Refinement of powder x-ray diffraction pattern at room temperature. Inset is a photograph of NCTO single crystal.}
\label{Structure}
\end{figure}

Na$_5$Co$_{15.5}$Te$_6$O$_{36}$ (NCTO) crystallizes in a hexagonal structure with space group $P$6$_3$/$m$ \cite{Na5Co15.5Te6O36-arXiv,Na5Co15.5Te6O36-JSSC-211-63}. As illustrated in Figs. \ref{Structure}(a-c), there are three crystallographic Co positions which are confirmed to be divalent by a Co $K$-edge x-ray absorption spectroscopy measurement \cite{Na5Co15.5Te6O36-arXiv}. Thereinto, the edge-shared Co(1)O$_6$ octahedra build zig-zag chains along the $c$ direction and are further connected to a kagome lattice in the $ab$ plane, whereas Co(2)O$_6$ and Co(3)O$_6$ trigonal prisms form face-shared chains in an alternate manner and sit in the center of each triangle. It is notable that Co3 site is partially occupied by Na atoms, giving rise to an on-site disorder. In view of the Ising anisotropy \cite{Na5Co15.5Te6O36-JSSC-211-63}, the topology of Co1 ions can be regarded as a stacked kagome lattice, see Fig. \ref{Structure}(d), and NCTO is likely the first example for $S$ = 1/2 stacked Ising KAF. In this work, we successfully grow NCTO single crystals and in detail study the magnetic ground state and peculiar field-induced transitions. Three successive magnetic transitions are detected at $T_1$ = 45.2 K, $T_2$ = 35.4 K, $T_3 \sim$ 20 K, respectively. When the magnetic field is applied along the Ising $c$ axis, a strikingly anomalous initial magnetization is unveiled below $T_3$. These exceptional phenomena indicate a possible partially disordered antiferromagnetic (PDA) state below $T_1$ and a frozen PDA state below $T_3$.

\begin{figure}
\includegraphics[clip,width=6.5cm]{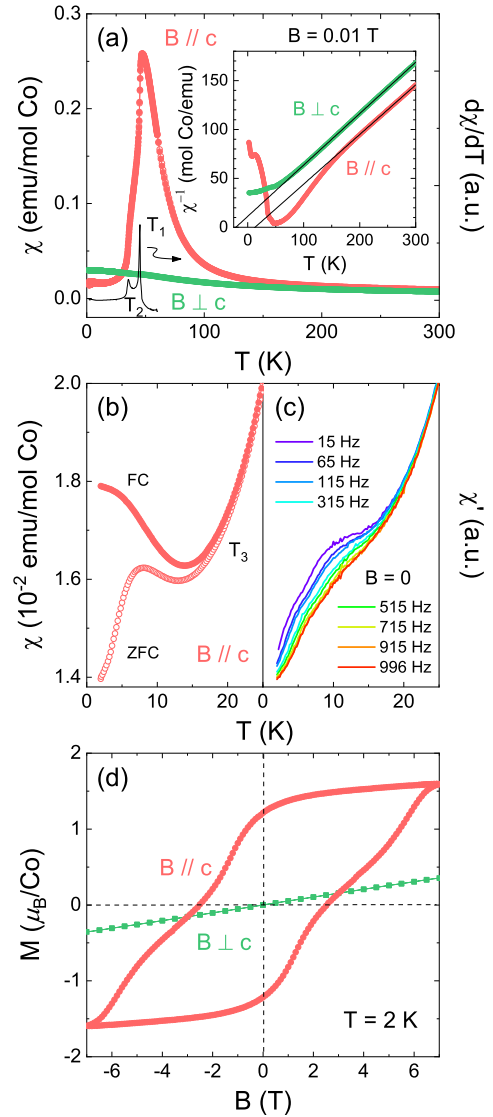}
\caption{(Color online) Anisotropic magnetic properties of NCTO. (a) Temperature dependencies of DC magnetic susceptibility $\chi_\parallel$ and $\chi_\perp$ for both field-cooling (FC, solid symbols) and zero-field-cooling (ZFC, open symbols) measurements in 0.01 T. The differential curve for $\chi_\parallel$ is also plotted to define the transition temperature $T_1$ and $T_2$. Inset are the reciprocals of $\chi_\parallel$ and $\chi_\perp$ measured after ZFC. Solid lines are the Curie-Weiss fittings. (b) A magnified view of $\chi_\parallel$ below 25 K. (c) AC susceptibility measured at different frequencies without an external magnetic field. The applied oscillation field is 2 Oe. (d) Magnetization $M_\parallel$ and $M_\perp$ at 2 K.}
\label{Ising}
\end{figure}

\section{Experiments}

NCTO single crystals were synthesized using a self-flux method. Na$_2$CO$_3$, Co$_3$O$_4$, and TeO$_2$ were used as starting materials with a molar ratio of 3:2.33:6. The mixture was ground thoroughly with alcohol and placed in an alumina crucible. The growth ampoule was then heated to 1000$^\circ$C in 10 hours, kept for 2 days, and slowly cooled to 600$^\circ$C with a rate of 2$^\circ$C/hr. After that, the furnace was cooled to room temperature in 7 hours. The as-grown single crystals shown in Fig. \ref{Structure}(e) are dark purple with the growth direction parallel to the $c$ axis. Phase purity was checked by the powder x-ray diffraction, see Fig. \ref{Structure}(e).

DC magnetic susceptibility was measured using one piece of crystal with mass of 21.7 mg between 2 and 300 K by using a SQUID-VSM (Quantum Design). AC magnetic susceptibility was performed using the same piece of crystal on the SQUID-VSM (Quantum Design) with applied frequencies between 15 Hz and 1000 Hz at the temperature range from 2 to 60 K. Heat capacity measurement was carried out on another piece of crystal about 9 mg using the relaxation method at the temperature range from 2 to 200 K on a PPMS (Quantum Design).

\section{Results and Discussions}

\subsection{Magnetic Anisotropy}

Figure \ref{Ising}(a) shows the temperature dependencies of the magnetic susceptibility with applied magnetic fields parallel ($\chi_\parallel$) and perpendicular ($\chi_\perp$) to the crystallographic $c$ direction. The large difference between $\chi_\parallel$ and $\chi_\perp$ reflects a strong Ising anisotropy with the easy axis along the $c$ direction. Upon cooling, $\chi_\parallel$ undergoes two successive AFM transitions, which are identified by the rapid drop at $T_1$ = 45.2 K and a shoulder-like anomaly at $T_2$ = 35.4 K. Both transitions are determined as the peak positions from the differential curve. With further lowering temperatures, a bifurcation between field-cooling (FC) and zero-field-cooling (ZFC) branches is observed below $T_3 \sim$ 20 K, see Fig. \ref{Ising}(b). Besides, a rather weak frequency dependence of AC susceptibility $\chi'$ is also found in Fig. \ref{Ising}(c). On the contrary, $\chi_\perp$ is monotonously increased with decreasing temperature and exhibits a very weak anomaly only around $T_1$. No ZFC/FC splitting is found in $\chi_\perp$. The high-temperature magnetic susceptibility behaves in a Curie-Weiss manner, and gives the Weiss temperature $\theta\rm_{CW}^{\parallel}$ = 13.1(9) K and $\theta\rm_{CW}^{\perp}$ = -21.1(2) K. The anisotropic Weiss temperature with a sign change suggests that the Curie-Weiss behavior is primarily originated from the single-ion anisotropy \cite{CM-12-966,PRB-95-094421,PRM-3-034005,PRB-93-104407,PRB-46-5425,PRM-3-074405,PRB-99-214402}. From the crystal structure, the $\angle$Co1-O-Co1 within the zig-zag chain is between 84.7$^\circ$ and 96.1$^\circ$, which favors a ferromagnetic (FM) intrachain coupling according to the Goodenough-Kanamori rule \cite{Goodenough}. The fast increase in $\chi_\parallel$ below 150 K is thus ascribed to the development of short-range FM correlation. Correspondingly, the transition at $T_1$ is associated with the AFM order of ferromagnetically coupled Co1 chains.

The distinct isothermal magnetization between parallel ($M_\parallel$) and perpendicular ($M_\perp$) to the $c$ direction in Fig. \ref{Ising}(d) confirms the strong single-axis anisotropy. The moment at 7 T is about 1.6 $\mu\rm_B$, consistent with the previous work \cite{Na5Co15.5Te6O36-JSSC-211-63}. The reduced moment of Co$^{2+}$ ions suggests an effective $S$ = 1/2 under the cooperative effect of spin-orbit coupling and crystal-field effect. The most remarkable feature is the presence of a large hysteresis loop in $M_\parallel$ up to 7 T. In view of the on-site disorder between Co3 and Na ions, the magnetic inhomogeneity is intuitively responsible for the observed ZFC/FC bifurcation in $\chi_\parallel$ and the frequency-dependent $\chi'$. However, this is not compatible with the pronounced magnetization hysteresis, ruling out the possibility of a glassy ground state. Considering the high fraction of Co1 ions $\sim$ 80\% in NCTO, the significant irreversible magnetization, which disappears above $T_3$ in Fig. S1 of Supplemental Material \cite{SI}, is probably dominated from the Co1 sublattice. The weak transition at $T_2$, however, is probably related to the AFM order of Co2/Co3 chains. This implies some weak but non-negligible interactions between adjacent Co1 and Co2/Co3 chains.

\subsection{Magnetic Transitions for $B \parallel c$}

Considering the Ising anisotropy of Co$^{2+}$ ions, below we will focus on the investigation of magnetic properties under the application of a magnetic field along the $c$ direction.

\begin{figure}
\includegraphics[clip,width=7.0cm]{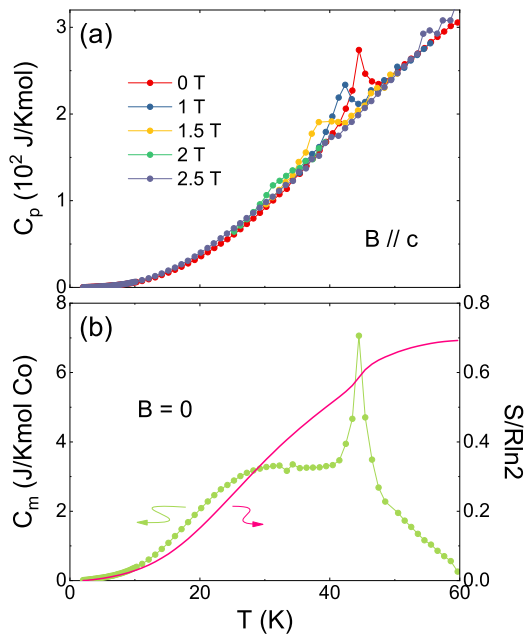}
\caption{(Color online) (a) Specific heat measured in magnetic fields along the $c$ axis up to 2.5 T. (b) Magnetic specific heat and estimated entropy recovery in zero field.}
\label{Cp}
\end{figure}

First, the specific heat $C\rm_p$ measured in various magnetic fields is shown in Fig. \ref{Cp}(a). It is of great interest to see that the zero-field $C\rm_p$ only displays one sharp $\lambda$ peak at $T_1$. At lower temperatures, $C\rm_p$ is smoothly decreased and there isn't any anomaly associated with the transition at $T_2$ or $T_3$. With increasing field, $T_1$ is gradually reduced and vanishes above $B^*$ = 2.5 T (Fig. S2). The temperature dependence of the zero-field $C\rm_p$ above 60 K can be reproduced using the Einstein model \cite{Tari} (Fig. S3), and gives two Einstein temperatures $\Theta\rm_{E1}$ = 199.3(6) K and $\Theta\rm_{E2}$ = 666(1) K. The magnetic specific heat is then extracted after subtracting the lattice contribution and the magnetic entropy is calculated to be 0.7$R$ln2, which is a little smaller than the expection for $S$ = 1/2 systems, as shown in Fig. \ref{Cp}(b).

\begin{figure}
\includegraphics[clip,width=7.5cm]{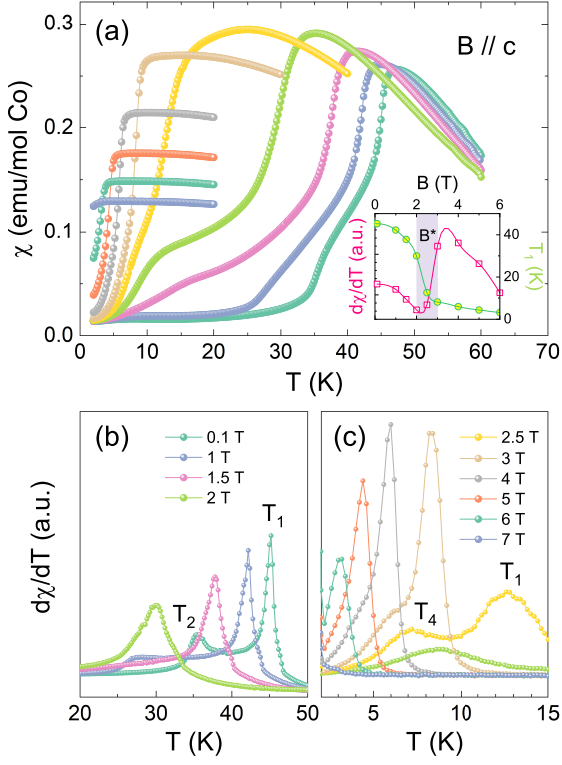}
\caption{(Color online) (a) Temperature dependencies of DC susceptibility measured in various fields after ZFC. The curve measured in 0.01 T, which is almost identical to the 0.1 T-one, is omitted for clarity. Inset shows the field dependencies of $T_1$ and the corresponding magnitude of the differential at $T_1$. $B^*$ is the critical field determined from the specific heat measurement. (b, c) The corresponding differential curves.}
\label{MT}
\end{figure}

Figure \ref{MT}(a) presents the temperature dependencies of $\chi_\parallel$ measured in various magnetic fields after ZFC. Although $T_1$ and $T_2$ are monotonously reduced with increasing field, $\chi_\parallel$ exhibits distinct tendency in low and high fields, as shown in Figs. \ref{MT}(b, c). On the one hand, $T_1$ is gradually reduced from 45.2 K in 0.1 T to 30 K in 2 T, and rapidly decreased to 12.6 K in 2.5 T; in higher fields $T_1$ is further reduced slowly again. On the other hand, d$\chi$/d$T$ at $T_1$ is firstly suppressed up to 2 T and then becomes enhanced again above $B^*$. Similarly, the feature at $T_2$ is firstly weakened with increasing field and nearly invisible in 1.5 T; above 2 T, another transition defined as $T_4$ appears and becomes strengthened above $B^*$. As field is increased further, the two transitions at $T_1$ and $T_4$ move closer and finally merge together in 4 T. In higher fields, this transition is continuously suppressed with a quickly weakened magnitude. The field dependencies of $T_1$ and the corresponding d$\chi$/d$T$ magnitude (inset to Fig. \ref{MT}(a)) indicate a clear anomaly around $B^*$, a critical field that completely suppresses the $\lambda$ peak in $C\rm_p$. It is noteworthy that contrary to the specific heat result, $T_1$ is persistent up to 6 T in $\chi_\parallel$.

\begin{figure}
\includegraphics[clip,width=8.5cm]{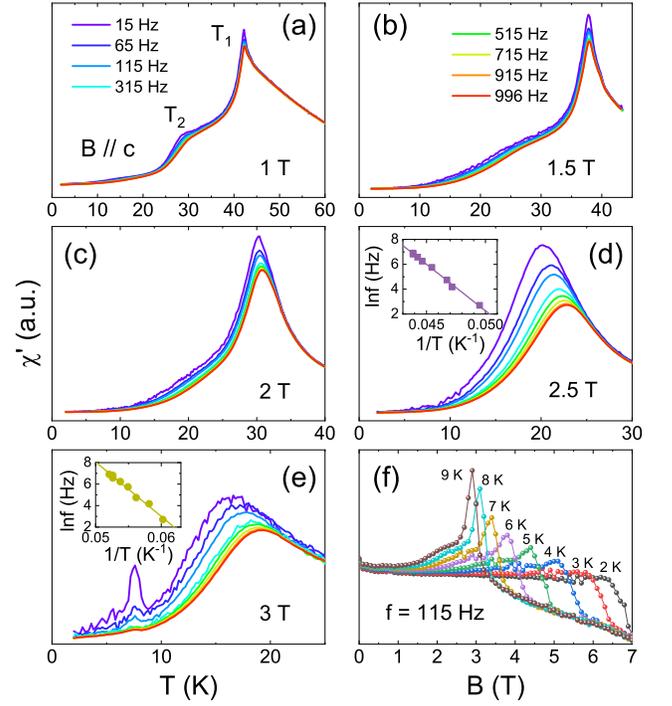}
\caption{(Color online) (a-e) Temperature dependencies of AC susceptibility measured with an oscillating field of 2 Oe. Insets in panels (d) and (e) are the fittings according to the Arrhenius law. (f) Field dependencies of AC susceptibility measured at different temperatures with $f$ = 115 Hz.}
\label{AC}
\end{figure}

AC susceptibility $\chi'$ as a function of temperature is carried out to further trace the magnetic transitions. In the absence of an external magnetic field, as shown in Fig. \ref{Ising}(c) and Fig. S4, $\chi'$ is independent of the frequency at $T_1$ and $T_2$ but a rather weak frequency dependence is found below $T_3$. As displayed in Figs. \ref{AC}(a-e), the transition at $T_2$ in 1 T also becomes frequency dependent, but $T_1$ is still robust except for a slightly reduced magnitude. The $T_1$-peak starts to show weak relaxation in 1.5 T which becomes obvious in 2 T. Meanwhile, the $T_2$- and $T_3$-transitions are too broad to distinguish. In $B^*$ = 2.5 T, a strong frequency-dependent peak is emergent around 20 K and the $T_1$-transition (12.6 K for $\chi_\parallel$ in 2.5 T) vanishes. According to the Arrhenius law, the derived energy barrier is about 748 K. This feature moves to lower temperatures in 3 T and the energy barrier is decreased to 507 K. In 3 T, the $T_1$-transition (8.2 K for $\chi_\parallel$) reappears again as evidenced by the presence of a weak anomaly. In higher fields, $T_1$ is further reduced though $\chi'$ is more noisy (Fig. S5).

\begin{figure}
\includegraphics[clip,width=7.5cm]{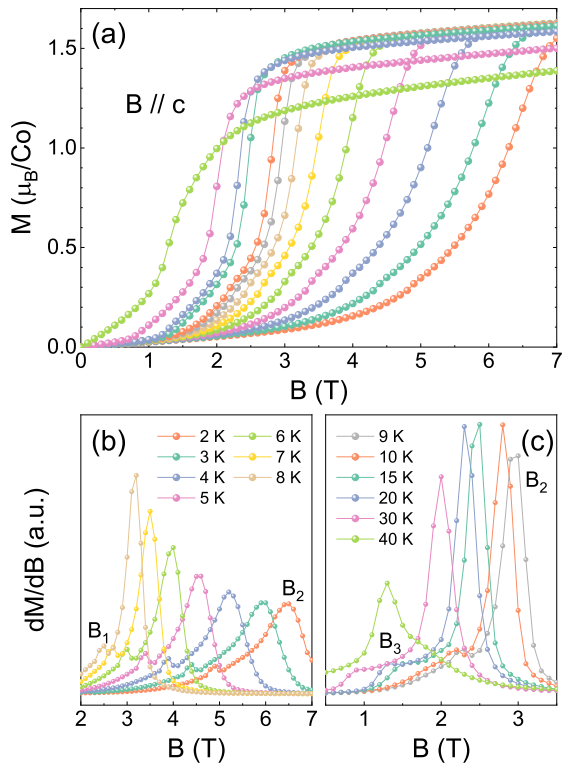}
\caption{(Color online) (a) Field dependencies of the initial magnetization in $M_\parallel$ measured at several constant temperatures after ZFC. (b, c) Differential curves below 8 K and above 9 K, respectively.}
\label{MH-virgin}
\end{figure}

Virgin magnetization measured at various temperatures are plotted in Fig. \ref{MH-virgin}. At 2 K, two magnetic transitions are detected at $B_1$ = 4.9 T and $B_2$ = 6.5 T, as determined from the differential curve in Fig. \ref{MH-virgin}(b). Upon warming, both $B_1$ and $B_2$ are decreased with a gradual enhancement of d$M$/d$B$ magnitude below 10 K. However, d$M$/d$B$ magnitude at $B_2$ keeps stable up to 20 K and then weakens at higher temperatures. At the meantime, the d$M$/d$B$ magnitude at $B_1$ which is evidenced as a small sharp peak at lower temperatures evolves into a broad shoulder above 15 K. Here we define $B_3$ as the critical field where the shoulder feature emerges. At 30 K, $B_3$ is about 0.9 T and becomes invisible at higher temperatures; $B_2$ is suppressed to 1.3 T at 40 K (Fig. S6). At 2 K, the magnetic entropy change in 7 T is calculated to be about 4.3 J/Kmol per Co, consistent with the specific heat result (Fig. S7).

\subsection{Anomalous Hysteresis Loop}

\begin{figure}
\includegraphics[clip,width=8.0cm]{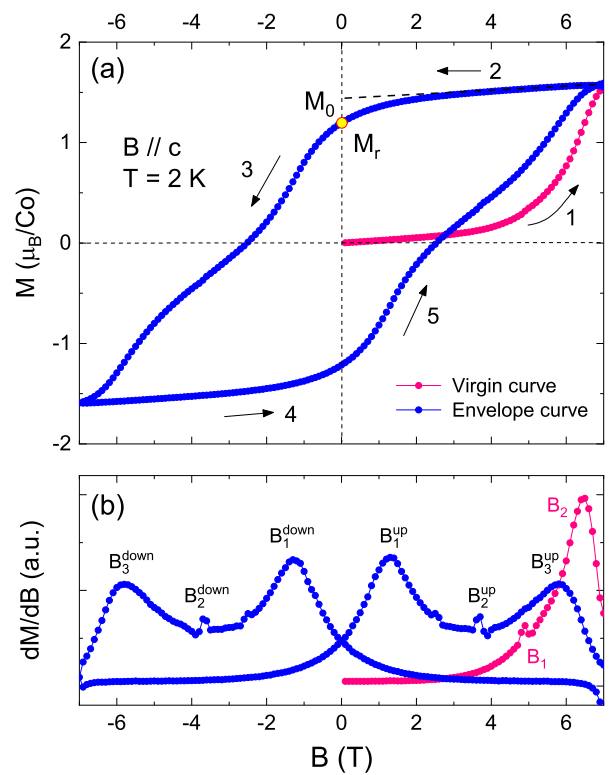}
\caption{(Color online) Magnetization of NCTO for $B \parallel c$ at $T$ = 2 K. (a) Magnetization hysteresis loop. The virgin curve is measured from 0 to 7 T after ZFC, followed by a complete field excursion (7 T $\rightarrow$ -7 T $\rightarrow$ 7 T). The numbered arrows denote the history of the field sweeping. The dashed line is a linear extrapolation from high fields with an intercept $M_0$. The remanent magnetization $M_r$ is represented by the yellow dot. (b) Corresponding differential curves. Multiple magnetic transitions are labeled by the critical fields.}
\label{MH-2K}
\end{figure}

When the field is applied along the $c$ axis, the magnetization presents a pronounced history dependence, and several interesting features are observed in Fig. \ref{MH-2K}. First, the virgin curve (path 1), which is recorded with ascending field from 0 to 7 T after cooling from the paramagnetic state in zero field, lies outside the hysteresis loop in higher fields. This is very different from conventional ferromagnets, in which the virgin curve locates within the loop due to the multiple domain effect. Second, the nearly linear increase of the moment in the low-field virgin curve is as expected for antiferromagnets. With further increasing field, two magnetic transitions are detected at $B_1$ and $B_2$. Third, when the field is released to zero (path 2), the disappearance of the magnetic transitions can be interpreted in terms of the domain reorientation in ferromagnets. Fourth, when the field is reversed (path 3), three successive magnetic transitions emerge at $B_1\rm^{down}$ = -1.3 T, $B_2\rm^{down}$ = -3.7 T, and $B_3\rm^{down}$ = -5.8 T (the weak anomaly at -2.5 T in Fig. \ref{MH-2K}(b) is due to the sign change of the magnetization). Fifth, in the second field-up process (path 5), the successive magnetic transitions appear again at $B_1\rm^{up}$ (= -$B_1\rm^{down}$), $B_2\rm^{up}$ (= -$B_2\rm^{down}$), and $B_3\rm^{up}$ (= -$B_3\rm^{down}$). It should be mentioned that the critical fields $B_2\rm^{up}$ and $B_3\rm^{up}$ in the subsequent field excursion are a bit lower than $B_1$ and $B_2$ in the virgin curve. Since the anomalous initial magnetization along with the higher critical fields is persistent up to $T_3$ (Fig. S1), a common origin should be in charge of both features.

This exceptional magnetization with concurrence of the large hysteresis loop and anomalous virgin curve is rarely observed in magnets. Two mechanisms from the literatures, metastable ground state and kinetic arrest of a first-order AFM-FM transition, may result in similar magnetization behavior. However, after careful comparisons with our experimental results, both mechanisms are considered to be not responsible for the anomalous magnetization hysteresis in NCTO. The reasons are as following:

\begin{enumerate}[(i)]
  \item After cooling in zero field, a metastable ground state could be developed due to the disordered domain walls. It has been reported in orthoferrite DyFeO$_3$ that the virgin magnetization curve shows two successive spin orientations below 2 K but only the higher-field one is kept in the following field excursion, giving rise to an irreversible initial magnetization \cite{PRB-89-224405}. Similar behavior is also observed in the field dependence of the capacitance in orthoferrite TbFeO$_3$ \cite{NatMater-11-694}. However, in both orthoferrites no large hysteresis loop is developed, in contrast to the significant hysteresis feature in NCTO. Besides, in the Y-type hexaferrite Ba$_{2-y}$Sr$_y$Co$_2$Fe$_{12-x}$Al$_x$O$_{22}$ ($x$ = 0.9), the initial magnetization is located outside the loop with a much lower magnitude, but the critical field is unchanged when the field is reversed \cite{PRB-101-075136}. This also disagrees with NCTO that the critical fields are higher in the virgin curve.
  \item Kinetics of a first-order transition, governed by time required to extract the latent heat of the system, can be arrested by rapid change of the environment. In some intermetallic alloys \cite{PRB-64-104416,PRB-73-020406,PRB-70-214421,PRB-97-134425}, the system has a AFM ground state and undergoes a field-induced AFM-FM transition. As the field is released, the kinetics of the FM-AFM transition is hindered with a fraction of frozen FM regime. The resultant state in zero field is thus a coexisting stable AFM phase along with the kinetically arrested metastable FM phase. When the field is ramped up again, the magnetization then starts from this coexistent state rather than the original AFM state as in the virgin curve. The forward magnetization is thus larger than the initial magnetization due to the larger moment of the FM phase, leaving the virgin curve lying below the hysteresis envelope. The kinetic arrest is always accompanied with a first-order transition, which is also contradictory to NCTO showing two second-order magnetic transitions at $B_1$ and $B_2$.
\end{enumerate}

\subsection{Phase Diagram}

\begin{figure}
\center\includegraphics[clip,width=8cm]{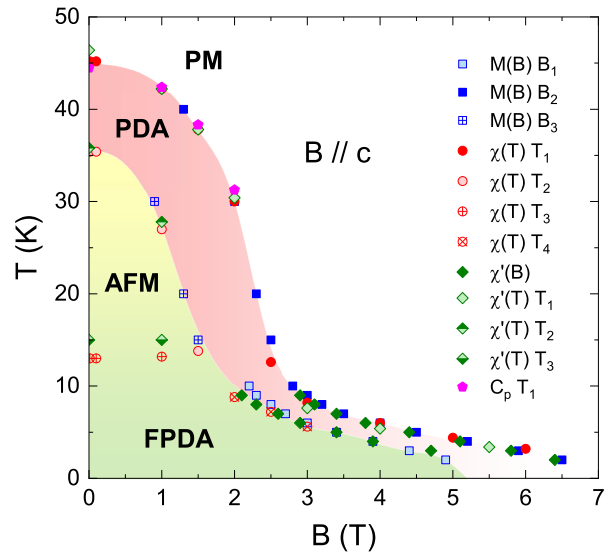}
\caption{(Color online) Magnetic field $vs$ temperature phase diagram of NCTO for $B \parallel c$. The data are extracted from the ZFC measurements through various experiments. PM, AFM, PDA, and FPDA denotes the paramagnetic phase, antiferromagnetic phase, partially disordered antiferromagnetic phase, and frozen partially disordered antiferromagnetic phase, respectively.}
\label{Phase-diagram}
\end{figure}

Based on the above experimental results, the magnetic field $vs$ temperature phase diagram for NCTO is illustrated in Fig. \ref{Phase-diagram}. As mentioned above, the transitions at $T_1$ and $T_3$ are probably associated with the Co1 sublattice, while the transition at $T_2$ is likely related to Co2/Co3 chains. The observed experimental phenomena below $T_3$, including the FC/ZFC bifurcation in $\chi_\parallel$, the spin-dynamic behavior in $\chi'$, and the pronounced magnetization hysteresis in $M_\parallel$, are reminiscent of the stacked TAFs $A_3BB'$O$_6$ ($A$ = Ca, Sr; $B$ = Co, Ni; $B’$ = Co, Rh, Ir). In this family, with descending temperature, the magnetic sublattice firstly develops a PDA state, in which 2/3 of the magnetically ordered chains align in opposite orientations and the rest 1/3 chains remain incoherent with each other and keep fluctuating. At lower temperatures, the incoherent 1/3 chains are freezed randomly (FPDA), resulting in a large ZFC/FC splitting in the magnetic susceptibility and a significant magnetization hysteresis; in addition, a slow spin relaxation behavior is uncovered by the AC susceptibility measurement \cite{PRB-65-180401(R),PRB-67-180404(R),EPJB-35-317,PRB-75-214422}. Nevertheless, it is noteworthy that in NCTO the specific heat shows a weak $\lambda$ anomaly at $T_1$ with 70\% entropy recovery. This is different from $A_3BB'$O$_6$, in which no signature associated with the PDA state is realized across the transition \cite{PRB-67-180404(R),PRB-75-214422,JPCM-15-5737,PRB-86-134409}, but similar to another stacked Ising TAF CsCoBr$_3$ with a $\lambda$ peak across the PDA state \cite{PRB-49-12299,PRB-12-5007}. Recently, an organic compound CoCl$_2$-2SC(NH$_2$)$_2$ with a triangular lattice is also reported to show a PDA ground state, in which the $\lambda$ anomaly is clearly observed in $C\rm_p$ with only 40\% expected entropy \cite{PRB-93-104407}.

For these reasons, the Co1 sublattice in NCTO is proposed to undergo a PDA state below $T_1$ and enter into a FPDA state below $T_3$. The transition at $T_2$ is then associated with the AFM order of Co2/Co3 chains. It should be mentioned that a recent neutron powder diffraction (NPD) measurement has proposed a canted AFM structure with a \textbf{k} = (0.5 0.5 0) propagation vector \cite{Na5Co15.5Te6O36-arXiv}, in which all the Co1 FM chains are arranged in a ferrimagnetic configuration on a basal triangle. However, the simple analysis of the NPD pattern at low temperatures is not informative enough to elucidate the nature of the magnetic order. As discussed in Ca$_3$CoRhO$_6$, both PDA and ferrimagnetic structures are explained qualitatively from NPD pattern \cite{JPSJ-70-1222}, but the PDA state is later confirmed through the powerful polarized neutron diffraction measurement \cite{PRL-87-177202}. To further examine the magnetic ground state and deeply understand the physics, local probes such as neutron scattering, nuclear magnetic resonance, and muon spin relaxation experiments are necessary to be carried out.

\subsection{PDA State in Kagome Lattice}

\begin{figure}
\includegraphics[clip,width=8.5cm]{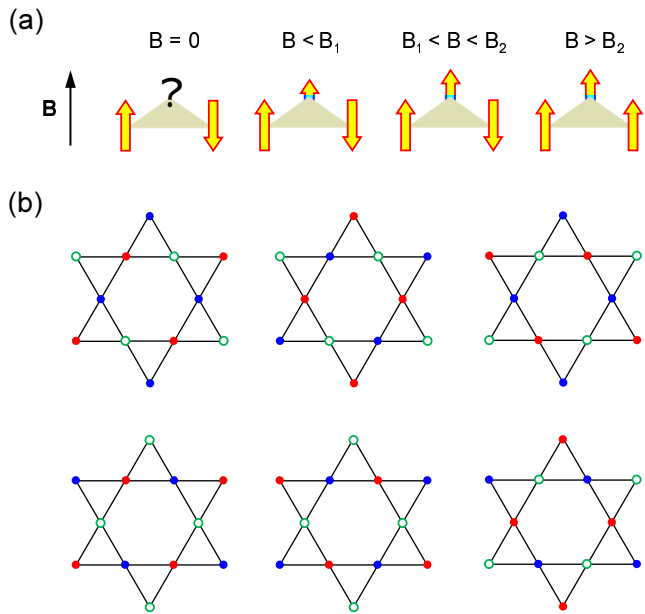}
\caption{(Color online) Proposed spin configurations in NCTO for $B \parallel c$. (a) Evolution of the spin structures of Co1 sublattice below $T_3$. The size of the arrow denotes the magnitude of Co1 moment. (b) Several configurations of the PDA state on a stacked kagome lattice. Green open circles denote the incoherent FM chains, and the red/blue solid circles denote the FM chains with up/down spins.}
\label{PDA}
\end{figure}

The topology of the Co1 sublattice can be viewed as a perfect $S$ = 1/2 stacked Ising kagome lattice, see Fig. \ref{Structure}(d). Since there are two field-induced magnetic transitions in the virgin curve, the proposed evolution of the spin configurations is sketched in Fig. \ref{PDA}(a). In zero field, 2/3 of the FM coupled Ising chains are anti-aligned and the rest 1/3 incoherent FM chains are freezed with random orientations. As the field is increased, the frozen chains are gradually directed to the field, which is responsible for the initially slow increase of the magnetization. At $B_1$, all the incoherent FM chains are oriented parallel to the magnetic field. With further increasing field, the FM chains with spins anti-parallel to the field flip, giving rise to the saturated magnetization at $B_2$.

Besides, the extensive degeneracy of the PDA state can also explain the higher critical fields in the virgin curve, as proposed in Sr$_3$NiIrO$_6$ \cite{PRB-94-224408}. Considering the coordination number ($z$ = 4) for a kagome lattice, each incoherent FM chain should be surrounded by two spin-up chains and two spin-down chains to ensure the cancellation of the AFM interchain interactions. Fig. \ref{PDA}(b) illustrates the six possible configurations of PDA state on a kagome lattice. When including further-neighbor interactions, novel $n$-sublattice PDA states might come out \cite{JPSJ-64-4609}. Since each incoherent FM chain can be surrounded by multiple configurations of ordered chains, the magnetic field should overcome the stochastic spread of the interchain interaction before the “chain flip” can be accomplished, which results in the higher critical fields in the virgin curve. In the subsequent field excursion, the degeneracy is then removed and the chain flips could be driven by a lower critical field.

From Fig. \ref{Phase-diagram}, it is clearly seen that the AFM state of Co2/Co3 chains is enclosed by the PDA and FPDA phases and is easily overcome above $B^*$ due to the rather weak interchain interactions. There is also a critical regime and several eccentric behaviors are found when the system passes through this area by either sweeping field around $B^*$ = 2.5 T or ramping temperature near $T^* \sim$ 15 K. First, the upper and lower boundaries of the PDA phase change to a much slower slope above $B^*$. Second, the $\lambda$ anomaly in $C\rm_p$ is smeared out above $B^*$. Third, $T_1$ determined from $\chi_\parallel$ shows a rapid drop around $B^*$, and meanwhile the d$\chi$/d$T$ magnitude exhibits a nonmonotonous field dependence. Fourth, the transition at $T_1$ in $\chi'$ disappears around $B^*$ and shows up again in higher fields. In particular, an obvious spin relaxation is observed in the vicinity of the critical field. Fifth, the d$M$/d$B$ magnitude follows a nonmonotonous temperature dependence near $T^*$. All these experimental observations imply a crossover behavior once the Co2/Co3 spins are fully polarized, which implies a non-negligible interplay between the Co1 sublattice and Co2/Co3 chains. In higher fields, the boundary between the PDA state and the paramagnetic state becomes undistinguishable. This is confirmed by the field dependence of the AC susceptibility with an applied $f$ = 115 Hz, as seen in Fig. \ref{AC}(f). At 2 K, $\chi'$ is nearly field independent in lower fields and exhibits a step-like decrease around 6.4 T. Upon warming, the decrease gradually transforms to a sharp peak with a gradually suppressed critical field. Besides, a weak anomaly located around 4.7 T is also found at 3 K and is developed to a shoulder centered at 2.1 T at 9 K.

\section{Conclusions}

In summary, we perform comprehensive characterizations on the magnetic ground state and field-induced transitions in Na$_5$Co$_{15.5}$Te$_6$O$_{36}$, and construct the magnetic field $vs$ temperature phase diagram. In this compound, zig-zag chains formed by the edge-shared Co(1)O$_6$ octahedra are arranged in a perfect kagome geometry, developing a stacked kagome lattice. This peculiar lattice is discussed to undergo a PDA state below $T_1$ = 45.2 K and enters into a FPDA state below $T_3 \sim$ 20 K. This scenario explains well the pronounced magnetization hysteresis and the exceptional initial magnetization lying outside the hysteresis loop with much higher critical fields. Na$_5$Co$_{15.5}$Te$_6$O$_{36}$ is likely the first example of $S$ = 1/2 stacked Ising KAF. This unusual composite geometry opens a brand-new field to search for $S$ = 1/2 Ising KAFs. Exploration with various Ising chains will promote the progress of novel phenomena and in-depth understanding of the relevant physics.

\begin{acknowledgements}

The work at Fujian Institute of Research on the Structure of Matter, CAS (Z.Y.Z., J.Y.L, and Z.Z.H.) was supported by the National Natural Science Foundation of China (Grants No. 52072368 and 21875249). The work at Anhui University (X.Y.Y.) was sponsored by the National Natural Science Foundation of China (Grants No. 12104010 and U1832209). The work at University of Science and Technology of China (N.L., H.L.C., and X.F.S.) was financed by the National Natural Science Foundation of China (Grants No. 11874336 and U1832209).

\end{acknowledgements}

\end{document}